\documentclass[a4paper]{article}
\usepackage{ASVspoof}
\usepackage{epsfig,amssymb,amsmath}
\usepackage{hyperref}
\usepackage{xcolor}
\usepackage{subcaption}
\usepackage{multirow}
\usepackage{balance}
\usepackage{verbatim}
\usepackage{lipsum}
\usepackage{fdsymbol}
\usepackage{booktabs}
\usepackage{mathrsfs}
\usepackage{cite}
\usepackage{hyperref}
\usepackage{cleveref}
\hypersetup{
    colorlinks=true,
    linkcolor=purple,
    filecolor=magenta,      
    urlcolor=purple,
}

\ninept
\hypersetup{nolinks=true}
\renewcommand{\vec}{\boldsymbol}

\title{ASVspoof~5: Crowdsourced Speech Data, Deepfakes, \\and Adversarial Attacks at Scale}

\makeatletter
\def\name#1{\gdef\@name{#1\\}}
\makeatother
\name{
{\em Xin Wang, H\'ector Delgado, Hemlata Tak, Jee-weon Jung, Hye-jin Shim, Massimiliano Todisco,} \\
{\em Ivan Kukanov, Xuechen Liu, Md Sahidullah,} \\
{\em Tomi Kinnunen, Nicholas Evans, Kong Aik Lee, Junichi Yamagishi} 
}      

\address{ASVspoof consortium \\ {\small \url{http://www.asvspoof.org/}}}
\begin{document}
\maketitle

\begin{abstract}

ASVspoof~5 is the fifth edition in a series of challenges that promote the study of speech spoofing and deepfake attacks, and the design of detection solutions.
Compared to previous challenges, the ASVspoof~5 database is built from crowdsourced data collected from a vastly greater number of speakers in diverse acoustic conditions.  
Attacks, also crowdsourced, are generated and tested using surrogate detection models, while adversarial attacks are incorporated for the first time.  New metrics support the evaluation of spoofing-robust automatic speaker verification (SASV) as well as stand-alone detection solutions, i.e., countermeasures without ASV. 
We describe the two challenge tracks, the new database, the evaluation metrics, baselines, and the evaluation platform, and present a summary of the results.  
Attacks significantly compromise the baseline systems, while submissions bring substantial improvements. 
\end{abstract}

\section{Introduction}
The ASVspoof initiative was conceived to foster progress in the development of detection solutions, also referred to as countermeasures (CMs) and presentation attack detection (PAD) solutions, to discriminate between bona fide and spoofed or deepfake speech utterances.
ASVspoof~5 is the fifth edition in a series of previously-biennial challenges \cite{Wu-ASVspoof2015,Kinnunen2017-assessing,asvspoof2019,yamagishiASVspoof2021Accelerating2021} and has evolved in the track definition, the database and spoofing attacks, and the evaluation metrics.

While the 2021 challenge edition involved distinct logical access (LA), physical access (PA), and speech deepfake (DF) sub-tasks~\cite{liuASVspoof2021Spoofed2023}, ASVspoof~5 takes the form of a single, combined LA and DF task, but encompasses two tracks: (i)~stand-alone spoofing and speech deepfake detection (CM, no ASV) and (ii)~spoofing-robust automatic speaker verification (SASV). 
Track 1 is similar to the DF track of the previous 2021 challenge. 
It reflects a scenario in which an attacker has access to the voice data of a targeted victim, e.g.\ data posted to social media. The attacker is assumed to use public data and speech deepfake technology to generate spoofed speech resembling the voice of the victim, and then to post the generated recordings to social media
, e.g. to defame the victim. Speech data, both bona fide and spoofed, may be compressed using conventional codecs (e.g., mp3) or contemporary neural codecs.

Track 2 shares the same goal as the LA sub-task of previous ASVspoof editions and the SASV2022 Challenge~\cite{sasv2022}. 
Track 2 assumes a telephony scenario where synthetic and converted speech are injected into a communication system (e.g.\ a telephone line) without any acoustic propagation. 
Participants can elect to develop single classifiers or separate, fused ASV and CM sub-systems. They can use either a pre-trained ASV sub-system provided by the organisers or can optimize their own bespoke system. 

Participants are furthermore provided with an entirely new ASVspoof~5 database. Source data and attacks, both crowdsourced, encompass greater acoustic variation than earlier ASVspoof databases.  The objectives are to evaluate the threat of spoofing and deepfake attacks forged using non-studio-quality data and optimised to compromise not just ASV sub-systems but also CM sub-systems. 
Source data, collected from a vastly greater number of speakers than for earlier ASVspoof databases, is extracted from the Multilingual Librispeech (MLS) English partition~\cite{pratap20_interspeech}.
In addition to the use of new spoofing attacks implemented using the latest text-to-speech (TTS) synthesis and voice conversion (VC) algorithms, adversarial attacks are introduced for the first time and combined with spoofing attacks.  

Also new is an \emph{open} condition for both tracks~1 and~2. In contrast to the traditional \emph{closed} condition, for which participants are restricted to use the specified data protocol, for the open condition participants have the opportunity to use external data and pre-trained speech foundation models, subject to there being no overlap between training data (i.e.~that used for training foundation models) and challenge evaluation data. 

A new suite of evaluation metrics is also introduced.
Inspired by the NIST SREs~\cite{NIST-SRE-CTSplan-2020}, we adopt the minimum detection cost function (minDCF) as the primary metric for Track 1. 
The log-likelihood-ratio cost function ($C_\text{llr}$) and actual DCF are also used to gauge not only discrimination but also calibration performance.
The recently proposed architecture-agnostic DCF (a-DCF)~\cite{shim2024dcf} is used as the primary metric for Track 2, with the tandem detection cost function (t-DCF)~\cite{Kinnunen2020-tDCF} and tandem equal error rate (t-EER)~\cite{Kinnunen2024-tEER} being complementary.

We present an overview of the two challenge tracks, the database, and the adopted metrics. Spoofing and deepfake attacks built by database contributors and their performance in fooling an ASV system are also described. Finally, we report a summary of system performance for the baselines and those submitted by 54 challenge participants.

\section{Database}
\label{sec:database}
The new ASVspoof~5 database has evolved in two aspects: source data and attack algorithms. In terms of the source data, it is built upon the MLS English dataset~\cite{pratap20_interspeech} to evaluate the performance of CM and SASV systems on detecting spoofing attacks forged using non-studio-quality data. The MLS English dataset incorporates data from more than 4k speakers, recorded with diverse devices. This is in contrast to the source database (VCTK~\cite{yamagishiCSTRVCTKCorpus2019}) in previous challenges, which contains around 100 speakers' data recorded in an anechoic chamber. The second major update of the ASVspoof~5 database is the stronger spoofing attacks. In addition to using the latest TTS and VC algorithms, the spoofing attacks are optimised to fool not only ASV but also CM surrogate sub-systems. This differs from previous ASVspoof challenge editions, where the organisers verified that spoofing attack data were successful in manipulating an ASV sub-system only. 
Based on the spoofing attacks, adversarial attacks are created using the Malafide \cite{panariello23b_interspeech} and Malacopula filters \cite{malacopula}. The former compromises the performance of CM, while the latter escalates the threat of the spoofed data to ASV.  
Last but not least, codecs, including a neural-network-based one, are applied to the bona fide and spoofed data.

\begin{table}[t!]
    \centering
    \setlength{\tabcolsep}{3pt}
    \caption{Summary of ASVspoof~5 database. The number of target speakers is listed in the bracket. Training and evaluation sets of track~1 do not define target speakers. Enrollment utterances are not counted. }
    \vspace{-3mm}
    \begin{tabular}{rlllll}
        \toprule
                  & \multicolumn{2}{c}{\#. speaker} & \multicolumn{2}{c}{\#. utterances} & \#. spf. \\
        \cmidrule(lr){2-3}\cmidrule(lr){4-5}
         & Female & Male & Bona fide & Spoofed & attack \\  
        \midrule
        Trn. & 196 & 204 & \phantom{0}18,797 & 163,560 & \phantom{0}8\\
        Dev. & 392 (196) & 393 (202) & \phantom{0}31,334 & 109,616 & \phantom{0}8\\
        Eva. T1 & 370 & 367  &  138,688 & 542,086 & 16 \\
        Eva. T2 & 370 (194) & 367 (173) & 100,708 & 395,924 & 16\\
        \bottomrule
    \end{tabular}
    \label{tab:database_statistics}
\end{table}

The database is built in three steps with the help of two groups of data contributors. 
First, the organisers partition the MLS English dataset into three disjoint subsets: A, B, and C. The data contributors of the first group use the MLS partition A and build TTS systems. With the spoofed data from those TTS systems, the organisers train surrogate ASV and CM systems (\S~\ref{sec:baselines}). 
In the second step, the data contributors of the second group use MLS partition B to build TTS and VC systems. They clone the target speakers' voices in the MLS partition B, query the surrogate ASV and CM systems to gauge the ``effectiveness" of the cloned voices, and tune their TTS and VC systems. 
Finally, the tuned TTS and VC systems are used to clone the target speakers' voices in the MLS partition C. Some of these TTS and VC systems are further combined with the adversarial attacking techniques (i.e., Malafide and Malacopula filters). 
Note that, to avoid potential data leakage, spoofing attacks and surrogate systems are built with privileged protocols, which are not shared with the challenge participants. 

The bona fide data of the ASVspoof~5 challenge training set is from the speakers in MLS partition A, and the spoofed data are from TTS systems built by the first data contributor group. The bona fide data of the development and evaluation sets are from MLS partition C, and the spoofed data are created by the second data contributor group. The speakers in the ASVspoof~5 challenge training, development, and evaluation sets are disjoint. The statistics are listed in Table~\ref{tab:database_statistics}.\footnote{MLS English dataset is scraped from the same source as Librispeech~\cite{panayotov2015librispeech}. Because challenge participants in the open condition may use models pre-trained on Librispeech, we remove speakers in the evaluation set who also appear in Librispeech.}

The spoofing attacks in training, development, and evaluation sets are also disjoint. Brief information about the attacks is listed in Table~\ref{tab:database_spoof_attacks}. 
In addition to classical TTS and VC algorithms (e.g., MaryTTS \cite{steiner2018creating}), the spoofing attacks in ASVspoof~5 use the latest DNN-based methods (e.g., ZMM-TTS \cite{gong2023zmm}). 
Two pre-trained systems, namely YourTTS~\cite{casanova2022yourtts} and XTTS~\cite{casanova2024xtts}, are used to clone target speakers' voices in a zero-shot manner.

\begin{table}[t!]
    \centering
    \setlength{\tabcolsep}{2pt}
    \caption{Summary of spoofing attacks. A01-A08, A09-A16, and A17-A32 are in training, development, and evaluation sets, respectively. AT denotes adversarial attack using Malafide, Malacopula, or both.}
    \vspace{-3mm}
    \resizebox{.5\textwidth}{!}{
    \begin{tabular}{cll|cll}
    \toprule
    ID & Type  & Algorithm & ID & Type & Algorithm  \\
    \toprule
    A01 & TTS & GlowTTS~\cite{kim2020glow}           
    &A17 & TTS & ZMM-TTS~\cite{gong2023zmm} \\
    A02 & TTS & variant of A01                       
    &A18 & AT & A17+Malafide \\
    A03 & TTS & variant of A01                       
    &A19 & TTS & MaryTTS \cite{steiner2018creating} \\
    A04 & TTS & GradTTS~\cite{popov2021grad}         
    &A20 & AT  & A12+Malafide \\
    A05 & TTS & variant of A04                       
    &A21 & TTS & A09+BigVGAN~\cite{lee2022bigvgan} \\
    A06 & TTS & variant of A04
    &A22 & TTS & varant of A09~\cite{luxExact2023} \\
    A07 & TTS & FastPitch~\cite{l2021fastpitch}      
    &A23 & AT  & A09+Malafide \\
    A08 & TTS & VITS~\cite{kim2021vits}              
    &A24 & VC  & In-house ASR-based \\ \cline{1-3}
    A09 & TTS & ToucanTTS~\cite{lux2023toucan}       
    &A25 & VC  & DiffVC~\cite{popov2021diffusion} \\
    A10 & TTS & A09+HifiGANv2~\cite{jungil20hifi}   
    &A26 & VC  & A16+original genuine noise \\
    A11 & TTS & Tacotron2~\cite{shen2018natural}     
    &A27 & AT  & A26+Malacopula \\
    A12 & TTS & In-house unit-select                 
    &A28 & TTS & Pre-trained YourTTS~\cite{casanova2022yourtts} \\
    A13 & VC  & StarGANv2-VC~\cite{li21estargan}     
    &A29 & TTS & Pre-trained XTTS~\cite{casanova2024xtts} \\
    A14 & TTS & YourTTS~\cite{casanova2022yourtts}   
    &A30 & AT  & A18+Malafide+Malacopula \\
    A15 & VC  & VAE-GAN~\cite{albadawy20vaegan}      
    &A31 & AT  & A22+Malacopula \\
    A16 & VC  & In-house ASR-based                   
    &A32 & AT  & A25+Malacopula \\
    \bottomrule
    \end{tabular}}
    \label{tab:database_spoof_attacks}
\end{table}

\begin{table}[t!]
    \centering
    \caption{Summary of codec and compression conditions in evaluation sets of Track 1 ($\largewhitestar$) and Track 2 ($\largeblackstar$).}
    \vspace{-3mm}
    \setlength{\tabcolsep}{4pt}
    \begin{tabular}{cllll}
        \toprule
         & Codec	&  Bandwidth  & Bitrate range & Usage \\ 
        \midrule
        C00 & - & 16 kHz & - & $\largewhitestar$ $\largeblackstar$ \\ 
        C01 & opus & 16 kHz & 6.0 - 30.0 & $\largewhitestar$ $\largeblackstar$ \\ 
        C02 & amr & 16 kHz & 6.6 - 23.05 & $\largewhitestar$ $\largeblackstar$\\ 
        C03 & speex & 16 kHz & 5.75 - 34.20 & $\largewhitestar$ $\largeblackstar$\\ 
        C04 & Encodec~\cite{de2023encodec} & 16 kHz & 1.5 - 24.0 & $\largewhitestar$ \\ 
        C05 & mp3 & 16 kHz & 45 - 256 & $\largewhitestar$ \\ 
        C06 & m4a & 16 kHz & 16 - 128 & $\largewhitestar$ \\ 
        C07 & mp3+Encodec & 16 kHz & varied & $\largewhitestar$ \\ 
        C08 & opus & 8 kHz & 4.0 - 20.0 & $\largewhitestar$ $\largeblackstar$\\
        C09 & arm & 8 kHz & 4.75 - 12.20 & $\largewhitestar$ $\largeblackstar$\\
        C10 & speex & 8 kHz & 3.95 - 24.60 & $\largewhitestar$ $\largeblackstar$\\
        C11 & varied & 8 kHz & varied & $\largewhitestar$ $\largeblackstar$\\
        \bottomrule
    \end{tabular}
    \vspace{-3mm}
    \label{tab:codecs}
\end{table}

To evaluate the CM and SASV systems' performance when both bona fide and spoofed data are (lossy) encoded or compressed, the evaluation sets contain data treated with codecs listed in Table~\ref{tab:codecs}.
C1-C7 operates with a 16kHz sampling rate, while C8-C11 operates in an 8kHz narrow band setting. To create the narrow band data, bona fide and spoofed data are down-sampled to 8 kHz, processed with the codec, and up-sampled to 16 kHz. Condition C0 replicates the scenario without encoding or compression. Bona fide and spoofed utterances are treated with one of the codec conditions. All the data are saved in a FLAC format with a sampling rate of 16 kHz. 
The leading and trailing non-speech segments in the evaluation set utterances have been removed. 

Participants of the closed condition of both Track 1 and 2 are required to use the same training and development sets to build their systems. For both tracks, participants in the open conditions can use external training data given the condition that it doesoverlapthe challenge database. They can use pre-trained speech foundation models built on some publicly available databases \cite[\S 4.2]{ASVspoof5_evalplan_phase2}.
The evaluation set for the two tracks covers the same set of utterances, except that Track 2 ignores a few codec conditions listed in Table~\ref{tab:codecs}.

\section{Performance measures}
This section provides a brief summary of the performance measures used in the two challenge tracks. 

\subsection{Track 1: from EER to DCF}

Track 1 submissions assign a real-valued bona fide-spoof detection score to each utterance. Different from past ASVspoof challenge editions for which EER was used as the primary metric for the comparison of spoofing CMs, Track 1 builds upon a \emph{normalized detection cost function} (DCF) \cite{NIST-SRE-CTSplan-2020}. While further details are available in \cite[Appendix]{ASVspoof5_evalplan_phase2}, the DCF has a simple form: 
    \begin{equation}\label{eq:dcf-track1-norm}
        \text{DCF}(\tau_\text{cm}) = \beta \cdot P_\text{miss}^\text{cm}(\tau_\text{cm}) + P_\text{fa}^\text{cm}(\tau_\text{cm}),
    \end{equation}
where $P_\text{miss}^\text{cm}(\tau_\text{cm})$ is the miss rate (false rejection rate of bona fide utterances) and $P_\text{fa}^\text{cm}(\tau_\text{cm})$ is the false alarm rate (false acceptance rate of spoofed utterances). Both are functions of a detection threshold, $\tau_\text{cm}$, and the constant $\beta$ in \eqref{eq:dcf-track1-norm} is defined as
    \begin{equation}
        \beta = \frac{C_\text{miss}}{C_\text{fa}}\cdot \frac{1-\pi_\text{spf}}{\pi_\text{spf}}, 
    \end{equation}
where $C_\text{miss}$ and $C_\text{fa}$ are, respectively, the costs of miss and false alarm, and where $\pi_\text{spf}$ is asserted prior probability of spoofing attack.\footnote{Since we have only two classes, it follows that $1-\pi_\text{spf}$ is the asserted prior of the bona fide class.}
The scenario envisioned in Track 1 lays on the assumption that, compared to spoofed utterances, bona fide speech utterances are, in general, far more likely in practice
(low $\pi_\text{spf}$). But, when encountered but not detected, the relative cost is high. We set $C_\text{miss}=1$, $C_\text{fa}=10$, $\pi_\text{spf}=0.05$, which gives $\beta\approx 1.90$. 
   
The normalized DCF in \eqref{eq:dcf-track1-norm} is used to compute both the \emph{minimum} and \emph{actual} 
DCFs. The former is the primary metric of Track 1, defined as $\text{minDCF} = \min_{\tau_\text{cm}} \text{DCF}(\tau_\text{cm})$.  
The latter, $\text{actDCF}=\text{DCF}(\tau_\text{Bayes})$, is the DCF evaluated at a fixed threshold $\tau_\text{Bayes} = -\log(\beta)$ 
under the assumption that the detection scores can be interpreted as log-likelihood ratios (LLRs). Whereas minDCF measures performance using an `oracle' threshold (set based on ground-truth), actDCF measures the realised cost obtained by setting the threshold to $\tau_\text{Bayes}$~\cite{NIST-SRE-CTSplan-2020}. 
Note that this is meaningful only when the scores can be interpreted as calibrated LLRs \cite{Brummer2006-application-independent,ferrer_calibration_tutorial}. 
Similar to the past challenge editions, ASVspoof~5 did not \emph{require} participants to submit LLR scores---rather, it was \emph{encouraged} for the first time.\footnote{Participants could post-process their raw detection scores into LLRs using implementations such as~\cite{ferrer_calibration_tutorial} in order to reduce actDCF. Note, however, that any order-preserving score calibration 
does not affect the primary minDCF metric.} 

Another complementary metric, \emph{cost of log-likelihood ratios} ($C_\text{llr}$)~\cite{Brummer2006-application-independent}, was used to assess the quality of detection scores when interpreted as LLRs:
    \begin{equation}
        C_\text{llr} = \frac{1}{2\log2}\Bigg(\frac{1}{|\mathscr{B}|}\sum_{s_i \in \mathscr{B}} \log\big(1 + e^{-s_i}\big)+\frac{1}{|\mathscr{S}|}\sum_{s_j \in \mathscr{S}}\log\big(1 + e^{s_j}\big)\Bigg),
    \end{equation}
where $\mathscr{B}=\{s_i\}$ and $\mathscr{S}=\{s_j\}$ denote, respectively, the sets of bona fide and spoofed trial scores. The lower the $C_\text{llr}$, the better calibrated (and more discriminative) the scores are. 
In addition to minDCF, actDCF, and $C_\text{llr}$, EER is also reported.

\subsection{Track 2: from SASV-EER to a-DCF}

For Track 2, participants were allowed to submit either single real-valued SASV scores 
or a triplet of scores 
which, in addition to SASV scores, contains two additional sets of spoofing (CM sub-system) and speaker (ASV sub-system) detection scores. While the former applies to any model architecture which outputs a single detection score, the latter assumes specific tandem (cascade) architecture~\cite{Kinnunen2020-tDCF} consisting of two clearly-identified sub-systems intended to detect spoofing attacks and to verify the speaker, respectively. In the latter case, the final SASV score is formed by combining the outputs of the two sub-systems (e.g., embeddings or scores) 
using an arbitrary combination strategy designed by the participants.

For both types of submission, the SASV 
scores are used to compute the primary challenge metric. 
Track 2 takes a step forward from EER-based metrics used in the first SASV challenge \cite{sasv2022} to DCF-based metrics. 
Extending upon the two-class DCF \eqref{eq:dcf-track1-norm}, ~\cite{shim2024dcf} recently proposed \emph{normalized architecture-agnostic} detection cost function (a-DCF)~\cite{shim2024dcf}, defined as 
\begin{equation}
\begin{aligned}\label{eq}
\text{a-}\text{DCF}(\tau_\text{sasv}) = &\alpha P_\text{miss}^\text{sasv}(\tau_\text{sasv}) + (1-\gamma) P_\text{fa,non}^\text{sasv}(\tau_\text{sasv}) \\
&+ \gamma P_\text{fa,spf}^\text{sasv}(\tau_\text{sasv})
\end{aligned},
\end{equation}
where $P_\text{miss}^\text{sasv}(\tau_\text{sasv})$ is the ASV miss (target speaker false rejection) rate and where $P_\text{fa,non}^\text{sasv}(\tau_\text{sasv})$  and $P_\text{fa,spf}^\text{sasv}(\tau_\text{sasv})$ are the false alarm (false acceptance) rates for non-targets and spoofing attacks, respectively. All three error rates are functions of an SASV threshold $\tau_\text{sasv}$. The constants $\alpha$ and $\gamma$ are given by
    \begin{equation}
        \begin{aligned}
            \alpha &= \frac{C_\text{miss}\pi_\text{tar}}{C_\text{fa,non}\pi_\text{non}+C_\text{fa,spf}\pi_\text{spf}},\\
            \gamma &= \frac{C_\text{fa,spf}\pi_\text{spf}}{C_\text{fa,non}\pi_\text{non}+C_\text{fa,spf}\pi_\text{spf}},
        \end{aligned}
    \end{equation}
where 
$C_\text{miss}$, $C_\text{fa,non}$, and $C_\text{fa,spoof}$ are the costs of miss, falsely acceptance of non-target speaker, and false acceptance of spoofing attack. Moreover, $\pi_\text{tar}$, $\pi_\text{non}$, and $\pi_\text{spoof}$ are the asserted priors of targets, non-targets (zero-effort impostors), and spoofing attack. The assumptions are similar to those in Track 1. We set $\pi_\text{tar}=0.9405$, $\pi_\text{non}=0.0095$, $\pi_\text{spf}=0.05$, $C_\text{miss}=1$ and $C_\text{fa,non}=C_\text{fa,spf}=10$. This gives $\alpha \approx 1.58$ and $\gamma \approx 0.84$.
The primary metric of Track 2 is the minimum of the a-DCF, 
        $\text{min a-DCF} = \min_{\tau_\text{sasv}} \text{a-DCF}(\tau_\text{sasv})$. 

For the submissions that contain clearly-identified ASV and CM sub-systems, 
\emph{ASV-constrained minimum tandem detection cost function} (t-DCF)~\cite{Kinnunen2020-tDCF} and \emph{tandem equal error rate}~(t-EER) \cite{Kinnunen2024-tEER} metrics are additionally reported. 
Whereas the former has served as the primary metric since ASVspoof 2019, the latter provides a complementary parameter-free measure of class discrimination. To compute the t-DCF metric, we adopt the same costs and priors as above and use ASV scores produced by a common ASV system of the organiser in place of scores provided by participants. 
        This allows computation of the minimum `ASV-constrained' t-DCF in the same way as for the previous ASVspoof challenges and enables the comparison of different CM sub-systems when they are combined with a common ASV sub-system.

For computation of the t-EER metric, 
both the CM and ASV sub-system scores are used to obtain a single \emph{concurrent t-EER} value, denoted by $\text{t-EER}_\times$. It has a simple interpretation as the error rate at a unique \emph{pair} of ASV and CM thresholds, $\vec{\tau}^\times :=(\tau_\text{asv}^\times,\tau_\text{cm}^\times)$, at which the miss rate and the two types of false alarm rates (one for spoofing attacks, the other for non-targets) are equal: $\text{t-EER}_\times= P_\text{miss}^\text{tdm}(\vec{\tau}^\times) = P_\text{fa,non}^\text{tdm}(\vec{\tau}^\times) = P_\text{fa,spoof}^\text{tdm}(\vec{\tau}^\times)$. 
        The superscript `tdm' is used to emphasize the assumed tandem architecture.
        The t-EER can be seen as a generalisation of the conventional two-class, single system EER which provides an application-agnostic discrimination measure.
        


\begin{figure}[t]
\centering
\includegraphics[width=0.45\textwidth]{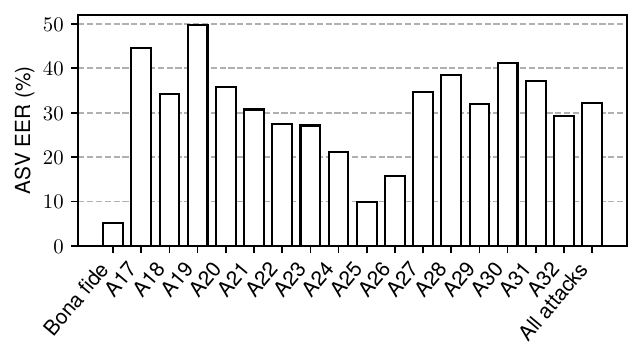}
\vspace{-5mm}
\caption{ASV EER  of organiser's common ASV on evaluation data. Results are pooled over all codec conditions. }
\label{fig:asv_result}
\end{figure}

\section{Common ASV, surrogate systems, and challenge baselines}
\label{sec:baselines}

\subsection{Common ASV system by organisers}
The common ASV system uses an ECAPA-TDNN speaker encoder~\cite{desplanques2020ecapa} and cosine similarity scoring. The ECAPA-TDNN model is trained using the training partitions of VoxCeleb 1 and 2~\cite{nagrani2017voxceleb}. 
The computed ASV cosine scores are subsequently normalised using an s-norm. 
Figure~\ref{fig:asv_result} illustrates the ASV EER values on the evaluation data. 
The EER (5\%) is low when discriminating between bona fide target and non-target speakers' data (the leftmost bar). However, the EERs are much higher when spoofing attacks are mounted. Note that although A25 is the least effective attack, it is more threatening when enhanced as an adversarial attack A32. 


\subsection{Baseline systems}
Track 1 adopts two CM baseline systems: RawNet2~\cite{tak2021rawnet} (B01) and AASIST~\cite{jung2022aasist} (B02). Both systems are end-to-end systems operating directly on raw waveforms.
Inputs to these baseline systems are raw waveforms of 4 seconds duration (64,000 samples). RawNet2 is composed of a fixed bank of $20$ sinc filters, six residual blocks followed by gated recurrent units, which convert the frame-level representation to utterance-level representation. The CM output scores are generated using fully-connected layers. 

AASIST uses a RawNet2-based encoder~\cite{tak2021rawnet} to extract spectro-temporal features from the input waveform. A spectro-temporal heterogeneous graph attention layers and max graph operations are then used to integrate temporal and spectral representations. CM output scores are generated using a readout operation and a linear output layer. 
Both baselines were trained with a weighted cross-entropy loss  for binary classification. 

In Track 2, a fusion-based~\cite{sasv2022} (B03) and a single integrated~\cite{mun2023towards} (B04) systems are adopted. B03 is adopted from the SASV 2022 challenge baseline but fuses the common ASV and the baseline AASIST of Track 1 using an LLR-based fusion tool~\cite{wangRevisiting2024}.
B04, which is based on MFA-Conformer~\cite{zhang2022mfa},  extracts a single embedding from the input waveform and produces a single SASV score. It is trained in three stages: speaker classification-based pre-training, copy synthesis~\cite{wang2023spoofed} training with adapted SASV loss functions, and in-domain fine-tuning.
VoxCeleb and copy synthesis data are used in the first and second stages, respectively. The in-domain fine-tuning is conducted using ASVspoof~5 training data.
The source codes for all baselines are accessible from the ASVspoof~5 git repository.\footnote{\url{github.com/asvspoof-challenge/asvspoof5}} 

\subsection{Surrogate systems}
The surrogate ASV system is based on ECAPA-TDNN and a probabilistic linear discriminant analysis scoring backend~\cite{prince2007probabilistic}. The surrogate CM systems include AASIST, RawNet2, LCNNs with LFCC, all of which are trained on bona fide data from the MLS partition A and spoofed attacks created by the first group of data contributors (see \S~\ref{sec:database}). Note that the surrogate CMs do not see attacks in the development and evaluation sets. 

%
\begin{figure}[t]
\centering
\includegraphics[width=0.45\textwidth]{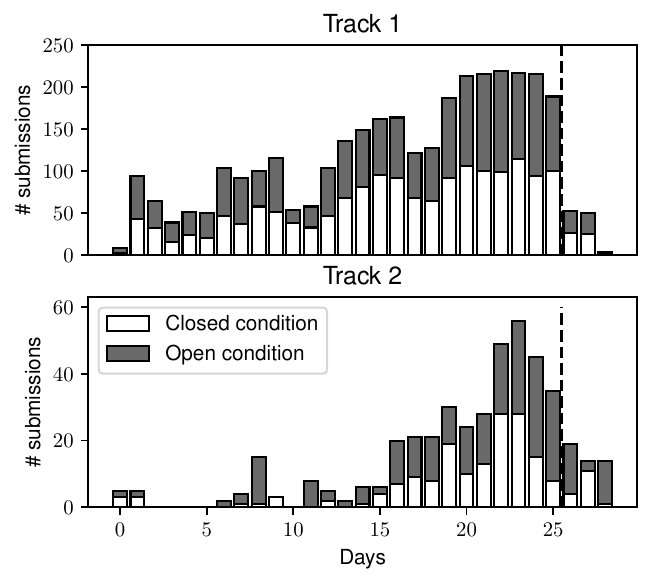}
\vspace{-3mm}
\caption{A \emph{stacked} bar chart showing the number of submissions to Track 1 and 2 for the Codalab \emph{progress} and \emph{evaluation} phases (last three days).}
\vspace{-3mm}
\label{fig:codalab_submission}
\end{figure}

\section{Evaluation platform}
ASVspoof~5 used the CodaLab website through which participants could submit detection scores and receive results.  The challenge was run in two phases, with an additional post-evaluation phase (not addressed in this paper).  
During the first \emph{progress} phase, participants could submit up to four submissions per day. Results determined from a subset of the evaluation set (i.e., the progress subset) were made available to participants who could opt to submit their results to an anonymised leaderboard. The evaluation phase ran for only a few days, during which participants could make only a single submission.  This submission was evaluated using the whole evaluation set. 

Figure~\ref{fig:codalab_submission} illustrates the number of submissions during the progress and evaluation phases. 
In Track 1, closed and open conditions received comparable amount of submissions.
In contrast, in case of Track 2, open condition received considerably higher number of submissions, demonstrating the need for additional data for training SASV systems.

\section{Challenge results}

\begin{table*}[!t]
\caption{Track 1 evaluation results. Submissions using a system ensemble and a single system are marked by $\bullet$ and $\circ$, respectively. Open-condition submissions using and not using pre-trained self-supervised models are marked by $\blacktriangle$ and $\triangle$, respectively. Submissions without providing a system description do not receive team IDs.
Submission after the initial deadline is \underline{underscored}.}
\label{tab:eval_results_t1}
\vspace{-5mm}
\begin{center}
\begin{tabular}{rrrrrr|rrrrrr}
\toprule
\multicolumn{12}{c}{\footnotesize{\textbf{Closed condition}}} \\ \midrule
\# & ID & minDCF & actDCF & $C_{\text{llr}}$ & EER & \# & ID & minDCF & actDCF & $C_{\text{llr}}$ & EER \\ \midrule
$\bullet$ $\phantom{0}1$ & T32 & 0.2436 & 0.9956 & 0.9458 & 8.61 & $\phantom{0}$ $18$ & - & 0.5990 & 0.9666 & 6.6313 & 24.12\\
$\bullet$ $\phantom{0}2$ & T47 & 0.2660 & 0.3380 & 0.6091 & 9.18 & $\phantom{0}$ $19$ & - & 0.6086 & 0.6091 & 0.8265 & 28.65\\
$\bullet$ $\phantom{0}3$ & T24 & 0.2975 & 0.2976 & 0.4182 & 10.43 & $\bullet$ $20$ & \underline{T07} & 0.6285 & 1.0000 & 1.0752 & 25.47\\
$\bullet$ $\phantom{0}4$ & T45 & 0.3948 & 1.0000 & 0.8515 & 14.33 & $\bullet$ $21$ & T27 & 0.6339 & 1.0937 & 1.0808 & 26.17\\
$\bullet$ $\phantom{0}5$ & T13 & 0.4025 & 0.4218 & 0.5238 & 14.75 & $\phantom{0}$ $22$ & - & 0.6463 & 0.8388 & 2.3251 & 26.45\\
$\phantom{0}$ $\phantom{0}6$ & - & 0.4079 & 0.4299 & 0.5512 & 14.16 & $\bullet$ $23$ & T41 & 0.6543 & 0.7641 & 0.9184 & 26.28\\
$\phantom{0}$ $\phantom{0}7$ & - & 0.4390 & 0.6332 & 0.8531 & 17.09 & $\circ$ $24$ & T06 & 0.6598 & 1.0000 & 1.1159 & 28.41\\
$\bullet$ $\phantom{0}8$ & T46 & 0.4783 & 1.0000 & 1.0509 & 20.45 & $\phantom{0}$ $25$ & - & 0.6617 & 0.9894 & 0.9562 & 27.31\\
$\bullet$ $\phantom{0}9$ & T23 & 0.5312 & 1.0000 & 1.1171 & 20.13 & $\circ$ $26$ & T14 & 0.6618 & 0.9307 & 2.4858 & 25.32\\
$\phantom{0}$ $10$ & - & 0.5340 & 1.0000 & 1.0228 & 19.10 & $\phantom{0}$ $27$ & - & 0.6989 & 0.7006 & 1.6935 & 31.15\\
$\phantom{0}$ $11$ & - & 0.5357 & 0.9533 & 3.3069 & 22.67 & $\circ$ $28$ & \textbf{{B02}} & 0.7106 & 0.9298 & 4.0014 & 29.12\\
$\bullet$ $12$ & T35 & 0.5505 & 1.0000 & 1.1435 & 23.42 & $\circ$ $29$ & T44 & 0.7997 & 1.0000 & 1.2774 & 35.15\\
$\phantom{0}$ $13$ & - & 0.5809 & 0.8537 & 4.0994 & 23.34 & $\phantom{0}$ $30$ & - & 0.8165 & 1.0000 & 1.1236 & 44.94\\
$\circ$ $14$ & T48 & 0.5813 & 0.9354 & 3.1923 & 23.63 & $\circ$ $31$ & \textbf{{B01}} & 0.8266 & 0.9922 & 4.0935 & 36.04\\
$\circ$ $15$ & T19 & 0.5891 & 0.6883 & 1.3277 & 24.59 & $\bullet$ $32$ & T54 & 0.8624 & 1.0000 & 1.1221 & 39.68\\
$\phantom{0}$ $16$ & - & 0.5895 & 1.0000 & 0.9351 & 23.93 & $\circ$ $33$ & T53 & 0.9744 & 1.0539 & 2.4977 & 44.94\\
$\phantom{0}$ $17$ & - & 0.5899 & 0.7470 & 1.3798 & 22.58\\
\bottomrule
\\
\toprule
\multicolumn{12}{c}{\footnotesize{\textbf{Open condition}}} \\ \midrule
\# & ID & minDCF & actDCF & $C_{\text{llr}}$ & EER & \# & ID & minDCF & actDCF & $C_{\text{llr}}$ & EER \\ \midrule
$\bullet$$\blacktriangle$ $\phantom{0}1$ & T45 & 0.0750 & 1.0000 & 0.7923 & 2.59 & $\phantom{0}$$\phantom{0}$ $18$ & - & 0.1949 & 0.2438 & 0.7028 & 7.05\\
$\bullet$$\blacktriangle$ $\phantom{0}2$ & T36 & 0.0936 & 1.0000 & 0.8874 & 3.41 & $\phantom{0}$$\phantom{0}$ $19$ & - & 0.1966 & 1.0000 & 0.9327 & 6.80\\
$\bullet$$\blacktriangle$ $\phantom{0}3$ & T27 & 0.0937 & 0.1375 & 0.1927 & 3.42 & $\bullet$$\blacktriangle$ $20$ & T33 & 0.2021 & 0.6028 & 0.5560 & 7.01\\
$\bullet$$\blacktriangle$ $\phantom{0}4$ & T23 & 0.1124 & 1.0000 & 0.9179 & 4.16 & $\phantom{0}$$\phantom{0}$ $21$ & - & 0.2148 & 1.0000 & 0.8124 & 7.43\\
$\bullet$$\blacktriangle$ $\phantom{0}5$ & T43 & 0.1149 & 0.5729 & 0.9562 & 4.04 & $\bullet$$\blacktriangle$ $22$ & T51 & 0.2236 & 1.0000 & 0.8011 & 7.72\\
$\bullet$$\blacktriangle$ $\phantom{0}6$ & T13 & 0.1301 & 0.1415 & 0.3791 & 4.50 & $\bullet$$\blacktriangle$ $23$ & T46 & 0.2245 & 1.0000 & 1.0308 & 9.36\\
$\bullet$$\blacktriangle$ $\phantom{0}7$ & T06 & 0.1348 & 0.2170 & 0.3096 & 5.02 & $\phantom{0}$$\phantom{0}$ $24$ & - & 0.2573 & 1.0000 & 0.9955 & 9.28\\
$\phantom{0}$$\phantom{0}$ $\phantom{0}8$ & - & 0.1414 & 0.5288 & 0.6149 & 4.89 & $\phantom{0}$$\phantom{0}$ $25$ & - & 0.2642 & 0.7037 & 2.1892 & 10.32\\
$\circ$$\blacktriangle$ $\phantom{0}9$ & T31 & 0.1499 & 0.2244 & 0.5559 & 5.56 & $\bullet$$\triangle$ $26$ & T47 & 0.2660 & 0.3321 & 0.4932 & 9.18\\
$\bullet$$\blacktriangle$ $10$ & T29 & 0.1549 & 0.2052 & 0.7288 & 5.37 & $\phantom{0}$$\phantom{0}$ $27$ & - & 0.2668 & 0.2923 & 0.6194 & 9.59\\
$\bullet$$\blacktriangle$ $11$ & T35 & 0.1611 & 1.0000 & 1.0384 & 5.93 & $\bullet$$\blacktriangle$ $28$ & T41 & 0.3010 & 0.3095 & 0.4773 & 10.45\\
$\phantom{0}$$\phantom{0}$ $12$ & - & 0.1665 & 0.1669 & 0.2351 & 5.77 & $\phantom{0}$$\phantom{0}$ $29$ & - & 0.4121 & 0.4266 & 0.7185 & 14.25\\
$\bullet$$\blacktriangle$ $13$ & T21 & 0.1728 & 0.2392 & 0.9498 & 6.01 & $\bullet$$\blacktriangle$ $30$ & T02 & 0.4845 & 1.0000 & 0.9332 & 17.08\\
$\circ$$\blacktriangle$ $14$ & T17 & 0.1729 & 1.0000 & 2.3217 & 5.99 & $\circ$$\triangle$ $31$ & T15 & 0.5112 & 0.6723 & 0.8858 & 22.24\\
$\circ$$\blacktriangle$ $15$ & T19 & 0.1743 & 0.3087 & 0.4757 & 6.06 & $\phantom{0}$$\phantom{0}$ $32$ & - & 0.6584 & 0.7451 & 1.1404 & 22.90\\
$\phantom{0}$$\phantom{0}$ $16$ & - & 0.1840 & 1.0000 & 0.8764 & 6.35 & $\phantom{0}$$\phantom{0}$ $33$ & - & 0.7969 & 1.0000 & 0.9920 & 35.72\\
$\phantom{0}$$\phantom{0}$ $17$ & - & 0.1933 & 1.0000 & 0.8342 & 6.67 & $\circ$$\triangle$ $34$ & T53 & 0.9744 & 1.0539 & 2.4977 & 44.94\\
\bottomrule
\end{tabular} 
\vspace{-4mm}
\end{center}
\end{table*}

\begin{table*}[!t]
\caption{Track 2 evaluation results. Submissions with only SASV scores are not evaluated using min t-DCF and t-EER. Submissions using a system ensemble and a single system are marked by $\bullet$ and $\circ$, respectively. Open-condition submissions using and not using pre-trained self-supervised models are marked by $\blacktriangle$ and $\triangle$, respectively. Submissions without providing a system description do not receive team IDs. Submission after the initial deadline is \underline{underscored}. REF is the organisers' ASV (\S~\ref{sec:baselines}) without a CM. }
\label{tab:eval_results_t2}
\vspace{-5mm}
\begin{center}
\begin{tabular}{rrrrr|rrrrr}
\toprule
\multicolumn{10}{c}{\footnotesize{\textbf{Closed condition}}} \\ \midrule
\# & ID & \shortstack{min \\ a-DCF} & \shortstack{min\\ t-DCF} & t-EER & \# & ID & \shortstack{min \\ a-DCF} & \shortstack{min\\ t-DCF} & t-EER\\ \midrule
$\bullet$ $\phantom{0}1$ & T45 & 0.2814 & - & - & $\bullet$ $\phantom{0}9$ & T23 & 0.4513 & 0.8279 & 49.34\\
$\bullet$ $\phantom{0}2$ & T24 & 0.2954 & 0.6175 & 9.58 & $\phantom{0}$ $10$ & - & 0.5130 & - & -\\
$\bullet$ $\phantom{0}3$ & T47 & 0.3173 & 0.5261 & 7.49 & $\circ$ $11$ & \textbf{{B04}} & 0.5741 & - & -\\
$\phantom{0}$ $\phantom{0}4$ & - & 0.3542 & - & - & $\phantom{0}$ $12$ & - & 0.6209 & 0.9073 & 25.39\\
$\phantom{0}$ $\phantom{0}5$ & - & 0.3744 & - & - & $\circ$ $13$ & \textbf{{B03}} & 0.6806 & 0.9295 & 28.78\\
$\phantom{0}$ $\phantom{0}6$ & - & 0.3893 & 0.7783 & 20.85 & $\circ$ $14$ & \textbf{{REF}} & 0.6869 & - & -\\
$\phantom{0}$ $\phantom{0}7$ & - & 0.3896 & - & - & $\phantom{0}$ $15$ & - & 0.8985 & - & -\\
$\phantom{0}$ $\phantom{0}8$ & - & 0.3971 & 0.7007 & 15.09\\
\bottomrule
\\
\toprule
\multicolumn{10}{c}{\footnotesize{\textbf{Open condition}}} \\ \midrule
\# & ID & \shortstack{min \\ a-DCF} & \shortstack{min\\ t-DCF} & t-EER & \# & ID & \shortstack{min \\ a-DCF} & \shortstack{min\\ t-DCF} & t-EER\\ \midrule
$\bullet$$\blacktriangle$ $\phantom{0}1$ & T45 & 0.0756 & - & - & $\phantom{0}$$\phantom{0}$ $\phantom{0}7$ & - & 0.1797 & 0.5430 & 8.39\\
$\bullet$$\blacktriangle$ $\phantom{0}2$ & T39 & 0.1156 & 0.4584 & 4.32 & $\phantom{0}$$\phantom{0}$ $\phantom{0}8$ & - & 0.3896 & - & -\\
$\bullet$$\blacktriangle$ $\phantom{0}3$ & T36 & 0.1203 & 0.4291 & 4.54 & $\phantom{0}$$\phantom{0}$ $\phantom{0}9$ & - & 0.4581 & - & -\\
$\bullet$$\blacktriangle$ $\phantom{0}4$ & T06 & 0.1295 & 0.4372 & 5.43 & $\circ$$\triangle$ $10$ & \textbf{{REF}} & 0.6869 & - & -\\
$\circ$$\blacktriangle$ $\phantom{0}5$ & \underline{T29} & 0.1410 & 0.4690 & 5.48 & $\phantom{0}$$\phantom{0}$ $11$ & - & 0.9134 & - & -\\
$\bullet$$\blacktriangle$ $\phantom{0}6$ & T23 & 0.1492 & 0.4075 & 4.63\\
\bottomrule 
\end{tabular} 
\vspace{-3mm}
\end{center}
\end{table*}

\subsection{Track 1}

Results on Track 1 are listed in Table~\ref{tab:eval_results_t1}. The baseline systems achieved minDCF higher than 0.7 and EERs higher than 29\%. Although they use the RawNet2 and AASIST architectures, which have been demonstrated to be effective on the previous ASVspoof challenge databases, the non-studio-quality data sourced from MLS and the more advanced spoofing attacks may have led to their unsatisfactory performance. 

It is encouraging that most of the submissions in the closed condition outperformed the baselines in terms of minDCF. The top-5 submissions succeed in obtaining minDCF values below 0.5 and EERs below 15\%, which is around 50\% relative improvement over the baselines. Similar to the trend observed in previous challenge editions, submissions using an ensemble of sub-systems tend to perform better. 

In the open condition, not surprisingly, the minDCF and EER values are lower than those in the closed condition. Notably, most of the well-performing submissions use features extracted by pre-trained self-supervised learning (SSL) models, e.g., wav2vec 2.0 (base version)~\cite{Baevski2020}. 

Despite the encouraging results, the top systems in both conditions obtained actDCF values close or equal to 1.0. The reason is that the systems' outputs are `normalized' to be between 0 and 1 rather than being calibrated to approximate LLRs. The scores are larger than the decision threshold specified by the priors and decision costs, which leads to $P_\text{miss}^\text{cm}(\tau_\text{cm})=0, P_\text{fa}^\text{cm}(\tau_\text{cm})=1$, and actDCF equal to 1.0. Similarly, their $C_\text{llr}$ values are not the best, suggesting again poor calibration. In contrast, some systems, such as T24 in the closed condition, are better calibrated. Although the primary metric is agnostic to score calibration, the top systems may consider further improvement via score calibration. 

\subsection{Track 2}

Results on Track 2 are listed in Table~\ref{tab:eval_results_t2}. Spoofing-robust ASV is technically more demanding than a stand-alone CM, which may be the reason for lower numbers of submissions to Track 2. 
B03 performed similarly to a reference system (REF) that is the same as B03 except using a random guessing CM sub-system. This indicates that the CM sub-system in B03 does not provide useful information for detection spoofing attacks. In contrast, the single integrated B04 performed better. However, note that the results on the baselines do not support the claim that fusion-based approach is inferior. In fact, all the top submissions are fusing ASV and CM sub-systems, including T45, which opted not to submit their ASV and CM scores.

Most of the submitted systems outperformed the baselines. Compared with the baseline, the top-3 submissions in the closed have 50\% relative improvement on the min a-DCF values. Similar to the findings in Track 1, submissions in the open condition reached lower metrics. The usage of SSL-based features is common among the top submissions.

\section{Conclusions}

This paper outlines the ASVspoof~5 challenge, which is designed to support the evaluation of both stand-alone speech spoofing and deepfake detection and SASV solutions. The fifth edition was considerably more complex than its predecessors, including not only a new task, but also more challenging crowdsourced data collected under variable conditions, 
spoofing attacks generated with a variety of contemporary algorithms optimised to fool surrogate ASV and CM sub-systems, and new adversarial attacks. 
Despite the use of lower-quality data to create spoofs and deepfakes,
detection performance for the baseline systems, all top-performing systems reported in recent years, is relatively poor. 
Encouragingly, results for most challenge submissions outperform the challenge baselines, sometimes by a substantial margin. 
We look forward to learning about the technical details from the challenge participants in their forthcoming research articles.
Results also reveal the hitherto ignored issue of score calibration, an essential consideration if detection solutions are deployed in real, practical scenarios.

With a particularly tight schedule for ASVspoof~5, more detailed analyses will be presented at the ASVspoof~5 workshop and reported in future work.

\section{Acknowledgements}

The ASVspoof~5 organising committee expresses its gratitude and appreciation to the challenge participants. For reasons of anonymity, they could not be identified in this article. Subject to the publication of their results and prior approval, they will be cited or otherwise acknowledged in future work.

The ASVspoof~5 organising committee extends its sincere gratitude to data contributors (in alphabetic order): Cheng Gong, Tianjin University; Chengzhe Sun, Shuwei Hou, Siwei Lyu, University at Buffalo, State University of New York; Florian Lux, University of Stuttgart; Ge Zhu, Neil Zhang, Yongyi Zang, University of Rochester; Guo Hanjie and Liping Chen, University of Science and Technology of China; Hengcheng Kuo 
 and Hung-yi Lee, National Taiwan University;
 Myeonghun Jeong, Seoul National University; Nicolas Muller, Fraunhofer AISEC; S\'ebastien Le Maguer, University of Helsinki; Soumi Maiti, Carnegie Mellon University;  Yihan Wu, Renmin University of China; Yu Tsao, Academia Sinica; Vishwanath Pratap Singh, University of Eastern Finland; Wangyou Zhang, Shanghai Jiaotong University.

The committee would like to acknowledge A$^\star$STAR (Singapore)  for sponsoring CodaLab platform, Pindrop (USA) and KLASS Engineering (Singapore) for sponsoring the ASVspoof 2024 Workshop. This work is also partially supported by JST, PRESTO Grant Number JPMJPR23P9, Japan and with funding received from the French Agence Nationale de la Recherche (ANR) via the BRUEL (ANR-22-CE39-0009) and COMPROMIS (ANR-22-PECY-0011) projects. This work was also partially supported by Academy of Finland (Decision No. 349605, project "SPEECHFAKES"). Part of this work used TSUBAME4.0 supercomputer at Tokyo Institute of Technology.

\bibliographystyle{IEEEbib}

%

\balance

\bibliography{ASVspoof_BibEntries}
\end{document}